\documentclass[preprintnumbers,amsmath,amssymb,aps, superscriptaddress,twocolumn]{revtex4-1}

\usepackage{graphicx}% Include figure files
\usepackage{dcolumn}% Align table columns on decimal point
\usepackage{bm}% bold math
\usepackage{natbib}
\usepackage{physics}
\usepackage[caption=false]{subfig}

\begin{document}

\title{Spin-Cooling of the Motion of a Trapped Diamond}

\author{T. Delord$^{*}$}
\author{P. Huillery$^{*}$}
\author{L. Nicolas}
\author{G. H\'etet} 

\affiliation{Laboratoire de Physique de l'Ecole normale sup\'erieure, ENS, Universit\'e PSL, CNRS, Sorbonne Universit\'e, Universit\'e Paris-Diderot, Sorbonne Paris Cit\'e, Paris, France.}

\begin{abstract}
Observing and controlling macroscopic quantum systems has long been a driving force in research on quantum physics. In this endeavor, strong coupling between individual quantum systems and mechanical oscillators is being actively pursued \cite{Treutlein2014, Rabl, Leibfried}. While both read-out of mechanical motion using coherent control of spin systems \cite{LaHaye, Kolkowitz, Arcizet, Lee_2017, Connell, Treutlein} and single spin read-out using pristine oscillators have been demonstrated \cite{MaminH, rugar}, 
temperature control of the motion of a macroscopic object using long-lived electronic spins has not been reported.
Here, we observe both a spin-dependent torque and spin-cooling of the motion of a trapped microdiamond.  Using a combination of microwave and laser excitation enables the spin of nitrogen-vacancy centers to act on the diamond orientation and to cool the diamond libration {\it via} a dynamical back-action. Further, driving the system in the non-linear regime, we demonstrate bistability and self-sustained coherent oscillations stimulated by the spin-mechanical coupling, which offers prospects for spin-driven generation of non-classical states of motion. Such a levitating diamond operated as a compass with controlled dissipation has implications in high-precision torque sensing \cite{Kim, Burgess, tongcangPRL}, emulation of the spin-boson problem \cite{Leggett} and probing of quantum phase transitions \cite{yin}. In the single spin limit \cite{Conangla} and employing ultra-pure nano-diamonds, it will allow quantum non-demolition read-out of the spin of nitrogen-vacancy centers under ambient conditions, deterministic entanglement between distant individual spins \cite{Rabl3} and matter-wave interferometry \cite{yin2013optomechanics, Scala, yin}.
\end{abstract}

\maketitle

%Since the observation of spin resonance using mechanical torque \cite{Alzetta}, 
Since the celebrated Einstein and de Haas' experiment in 1915 \cite{Einstein},
much work has been carried out in the detection of atomic spins through mechanical motion \cite{Alzetta}, culminating in the observation of a magnetic force from single spins \cite{rugar, MaminH} and magnetometry at the nanoscale \cite{Burgess}. 
Conversely, single spins and qubits have also been utilized to sense the motion of objects close to the quantum ground state.
Single qubit thermometry of mechanical oscillators at the quantum level was realized using superconducting qubit coupled to membranes \cite{LaHaye, Connell} and nitrogen-vacancy (NV) centers coupled to cantilevers \cite{Arcizet, Kolkowitz, Lee_2017}. 
A crucial next step is to reach strong coupling between long-lived spins and mechanical oscillators, which will enable ground state cooling, as in tethered quantum opto-mechanical platforms \cite{Arcizet2, Gigan, Schliesser}, and the observation of quantum superpositions of macroscopic systems \cite{Rabl}.
One further prospect is the entanglement between multiple spins \cite{Rabl3}, with far reaching implications in quantum information science and metrology \cite{Barson}. 
Obtaining coupling rates that surpass both the decoherence of the spin and of the mechanical system is however still a challenge for most state-of-the-art platforms.
Recently, there has been renewed focus on levitating objects \cite{Chang19012010, romero2010toward} motivated by the low mass and high Q-factors that they offer, together with the possibility to cool their motion using embedded spins \cite{yin2013optomechanics}. 
There is a strong analogy between this platform where spins move a levitating crystal and laser-cooled atoms where electrons move atomic nuclei. It may thus be forecast that a levitating particle containing a few long-lived spins will ultimately reach similar level of control as trapped ions \cite{Leibfried} with bright prospects for the aforementioned applications. 

\begin{figure*}[ht!]
\centerline{\scalebox{0.065}{\includegraphics{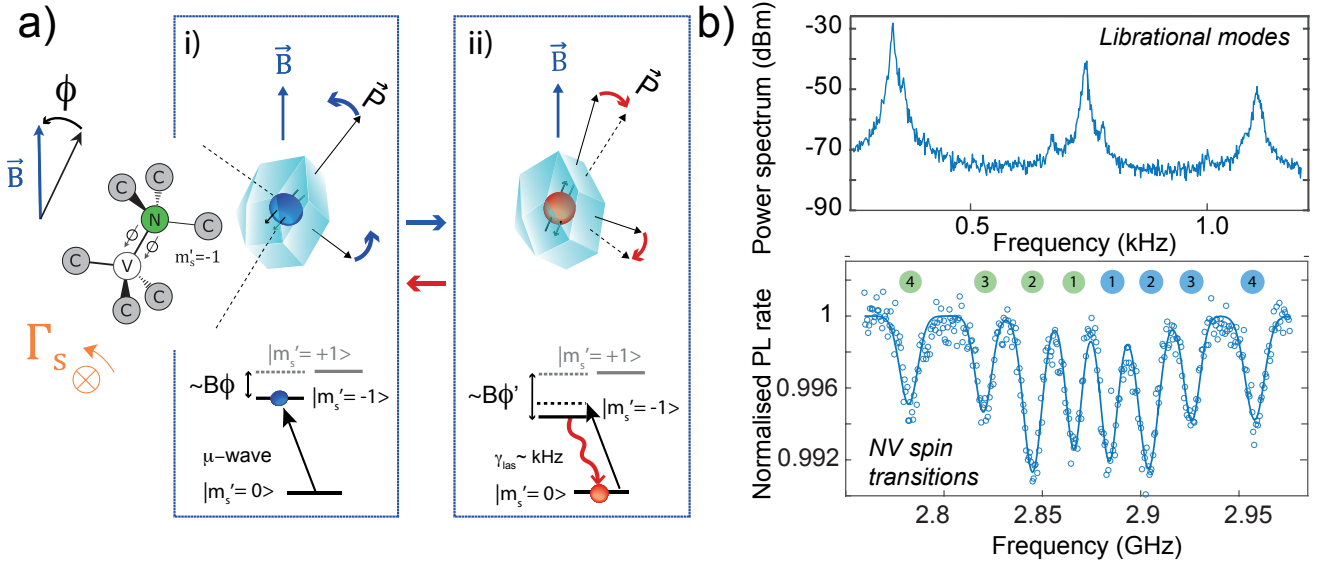}}}
\caption{{\bf Spin and mechanical systems}. a) Sketch of the diamond crystallographic structure hosting NV defects. i) Equilibrium position of the diamond in the Paul trap before the micro-wave excitation. The principal diamond axis $\vec{P}$ points towards the main trap axis.
The ground state spin-levels and microwave drive are depicted below. The magnetic field $\vec{B}$ lifts the degeneracy between the excited spin states by about $\gamma_e B\phi$, where $\phi$ is the angle between the magnetic field and the NV axis. The microwave signal prepares the NV in a magnetic state, which induces a torque to the diamond.  ii) Once at the new angular position, the spin projection onto the magnetic field is changed to $\gamma_e B \phi'$. The microwave is then no longer resonant : the spin relaxes to the ground state and the diamond returns to its initial position. 
b) Measurements of the three librational modes undergoing Brownian motion at 1~mbar of vacuum pressure (top) and of the typical electronic spin resonances from the NV ensemble within a microdiamond outside the trap using standard optically detected magnetic resonance (ODMR) at 30~G (below). Solid lines are a fit to the data.}\label{setup}
\end{figure*}

In this work, we report on a controllable torque induced by the spins of atoms embedded into a macro-object. Specifically, we couple the spin of many nitrogen-vacancy (NV) centers to the orientation of a trapped diamond particle. This coupling then enables us to show read-out of the NV centers spin resonance together with cooling and lasing of the diamond motion.

The crystallographic structure of NV centers is depicted in Fig.\ref{setup}-a). 
The spin-spin interaction between the two electrons in the NV center ground state lifts the degeneracy of the spin triplet eigenstates by $D= 2.87$~GHz at room temperature. 
Such an interaction implies that the NV center has a preferential quantization axis that is along one of the four crystal axes $\langle 111 \rangle$.
In the presence of a magnetic field ${\bf B}$ at an angle $\phi$ with respect to the NV axis, the energy difference between the two energy eigenstates $|m_s'=\pm 1\rangle$ is about $\gamma_e B \phi$, where $\gamma_e$ is the gyromagnetic ratio of the electron.
%Ignoring the influence of the strain and of the nuclear spins, the NV hamiltonian reads 
%$\hat{H}_{\rm NV}=\hbar D \hat{S}_z^2+ \hbar \gamma_e {\bf \hat{S}}\cdot{\bf B}$.
Spin control can then be performed using optical and microwave excitation and the angular dependence of the NV spin energy eigenstates is expected to allow rotation and cooling of the diamond angular motion.
%Optically-detected-magnetic-resonance (ODMR) and coherent spin control on angular stable diamonds can then be performed. 
%Another expected effect is the mechanical rotation and cooling of the diamond angular motion.
Once in a magnetic state {\it via} a resonant microwave excitation, the NV center will tend to align the corresponding diamond crystalline axis to the magnetic field, as illustrated in Fig.\ref{setup}-a-i). 
Further, laser triggered relaxation from the excited state can then extract the work exchanged between the spin magnetic energy and the librational motion (see Fig.\ref{setup}-a-ii)).

In our experiment, harmonic librational (sometimes called torsional, pendular or rotational) confinement is provided both by the Paul trap and the particle asymmetry. We measure the diamond libration using the laser reflection off the diamond surface. The $\mu$m size roughness on our 15 $\mu$m particle enables a specular pattern to be detected at the particle image plane, which after mode-matching one of the many bright spots to an optical fiber yields an angular sensitivity of about 0.3 mrad$/\sqrt{\rm Hz}$ and a resolution of about 10~mrad/Mcounts/s (see Methods).
Under vacuum conditions ($\approx$ 1 mbar), the signal power spectrum plotted in Fig.\ref{setup}-b) shows harmonic motion of the three librational modes with frequencies $\omega_\phi/2\pi$ ranging from 200~Hz to 1~kHz and with a damping rate of about 15~Hz. 
Fig.\ref{setup}-b) also shows an optically detected magnetic resonance (ODMR) spectrum for a diamond outside the trap, in the presence of a magnetic field $B\approx30$~G. Eight transitions, corresponding to the projections of the B field onto the four NV orientations are observed, with typical spin decoherence rates $1/T_2^*\approx 7$~MHz.

\begin{figure*}[ht!]
\centerline{\scalebox{0.058}{\includegraphics{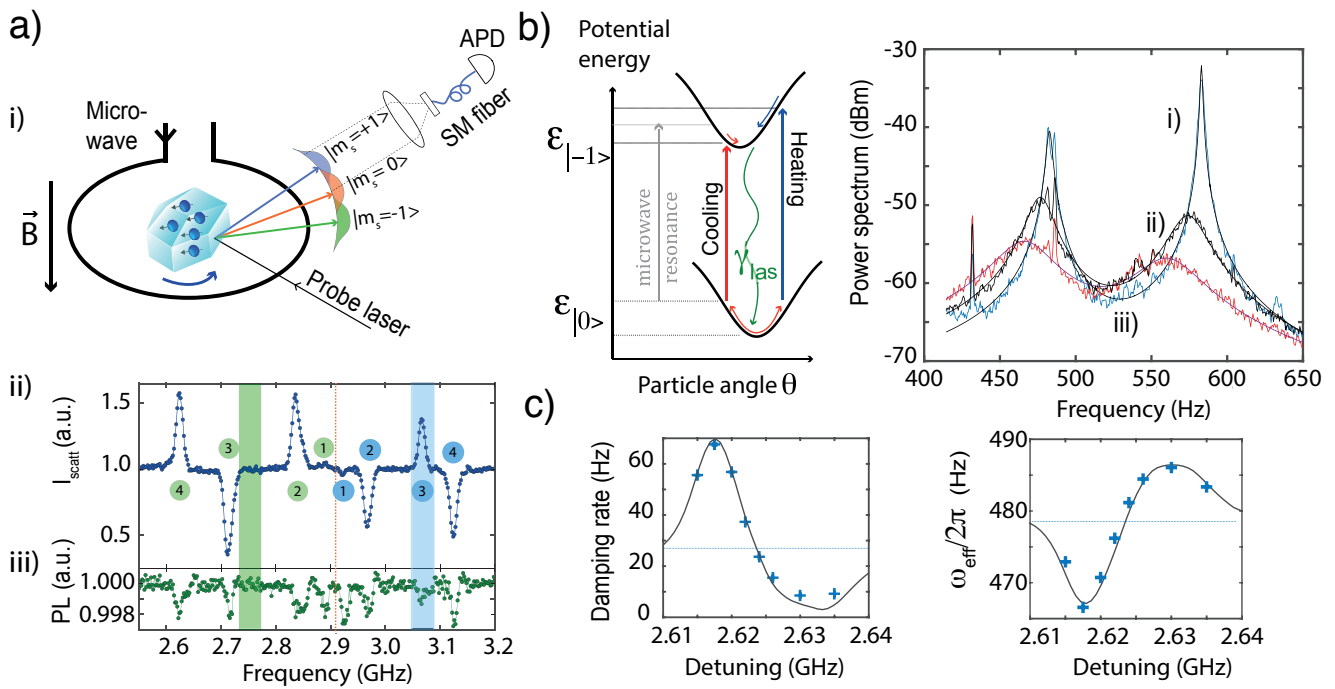}}}
\caption{{\bf Spin-dependent torque and cooling of a levitating diamond}  a) i) Sketch of the laser beam deflection induced by the NV spin-torque. ii) Detected APD count-rate $I_{\rm scatt}$ as a function of the microwave frequency. iii) corresponding ODMR.
b) Cooling/heating cycle of the librational motion induced by the spin-mechanical coupling (left). Power spectrum of the reflected light intensity from the diamond surface when the microwave is tuned to the blue (trace i)), to the center (trace ii)), and to the red (trace iii)) of the spin resonance (on the right). The alignment of the reflected light from the diamond surface in the fiber was optimized to only let these two librational modes appear in the power spectrum.  Note that the particle is here different from the one used in figure 1b. c) Damping rates and librational mode frequencies as a function of the microwave detuning. Plain lines show a fit to the experimental data using numerical simulations. 
}\label{Meca_SS}
\end{figure*}

We now measure the diamond rotation induced by the $N\approx 10^9$ NV electronic spins inside the diamond, with the same optical read-out as for the librational mode detection, as depicted in Fig. \ref{Meca_SS}-a)-i).  
The expected magnitude of the spin-torque is $\Gamma_s=\hbar N \gamma_e B S_z \approx 10^{-19}$N.m. Here $S_z$ is population in one of the magnetic states, determined by the competition between the microwave and laser polarization (both at rates in the 100 kHz range). This torque gives a displacement of the particle angle in the trap $\delta\phi=\Gamma_s/ I \omega_\phi^2 \approx 10$ mrad, where $I\approx 10^{-22}$~kg.m$^2$ is the particle moment of inertia.
As can be seen in Fig. \ref{Meca_SS}-a)-ii), sweeping a microwave around the spin resonances indeed enables conspicuous features to appear. Once in the magnetic state $|m_s'=-1\rangle$ or $|m_s'=+1\rangle$, the NV centers tend to align or anti-align the diamond orientation to the magnetic field, which is manifest in the anti-correlation between the detected intensity levels for all pairs of transitions. A standard ODMR is also measured under the same magnetic field amplitude and measurement time (see Fig. \ref{Meca_SS}-a)-iii)) demonstrating perfect correlation in the frequency of the peaks in the two measurements. 
%An upper bound for the observed angular displacements was obtained using NV magnetometry (see methods) and gives $\delta\phi\approx 40$ mrad, in good agreement with the above calculations. 

This spin-mechanical effect is in fact much richer than a static spin-dependent torque. As shown in Fig.\ref{setup}-a), the NV centers are magnetized through a microwave tone whose detuning from the NV resonances changes as the diamond rotates. 
To first order, such a torque will increase (resp. decrease) the confinement of the Paul trap if the microwave is blue (resp. red) detuned from the spin-resonance at the equilibrium angular position. Further, since the spin lifetime is on the order of the libration period, it enables dynamical back-action. A delay between the NV magnetization and the angular oscillation, observed in \cite{Huillery}, indeed induces a torque that depends on the velocity, in close analogy with opto-mechanical schemes \cite{Arcizet2, Gigan, Schliesser} and with Sisyphus cooling of cold atoms. 
The net result is a pronounced cooling (resp. heating) of the diamond motion when the microwave is red (resp. blue) detuned from the spin resonance as sketched in Fig.~\ref{Meca_SS}-b). In order to observe such spin-spring and spin-cooling effects, we monitor the librational power spectrum as a function of the microwave detuning from the electronic spin resonance. Fig.~\ref{Meca_SS}-b) shows the result of measurements taken for three different microwave frequencies with a Rabi frequency of 10 kHz. A strongly modified spring and damping of the mechanical mode are observed. Assuming that the initial temperature is 300~K (see Methods), the resulting temperature after spin-cooling is here 80~K. Fig.~\ref{Meca_SS}-c) shows measurements of the damping rate and spring effects as a function of microwave frequency in good agreement with a theoretical model (see Methods). Cooling is at present limited by heating from the microwave excitation of the motion on the blue side. This could be eliminated by increasing the trapping frequency $\omega_\phi/2\pi$ above the NV spin transition linewidth. 

\begin{figure*}[ht!]
\centerline{\scalebox{0.20}{\includegraphics{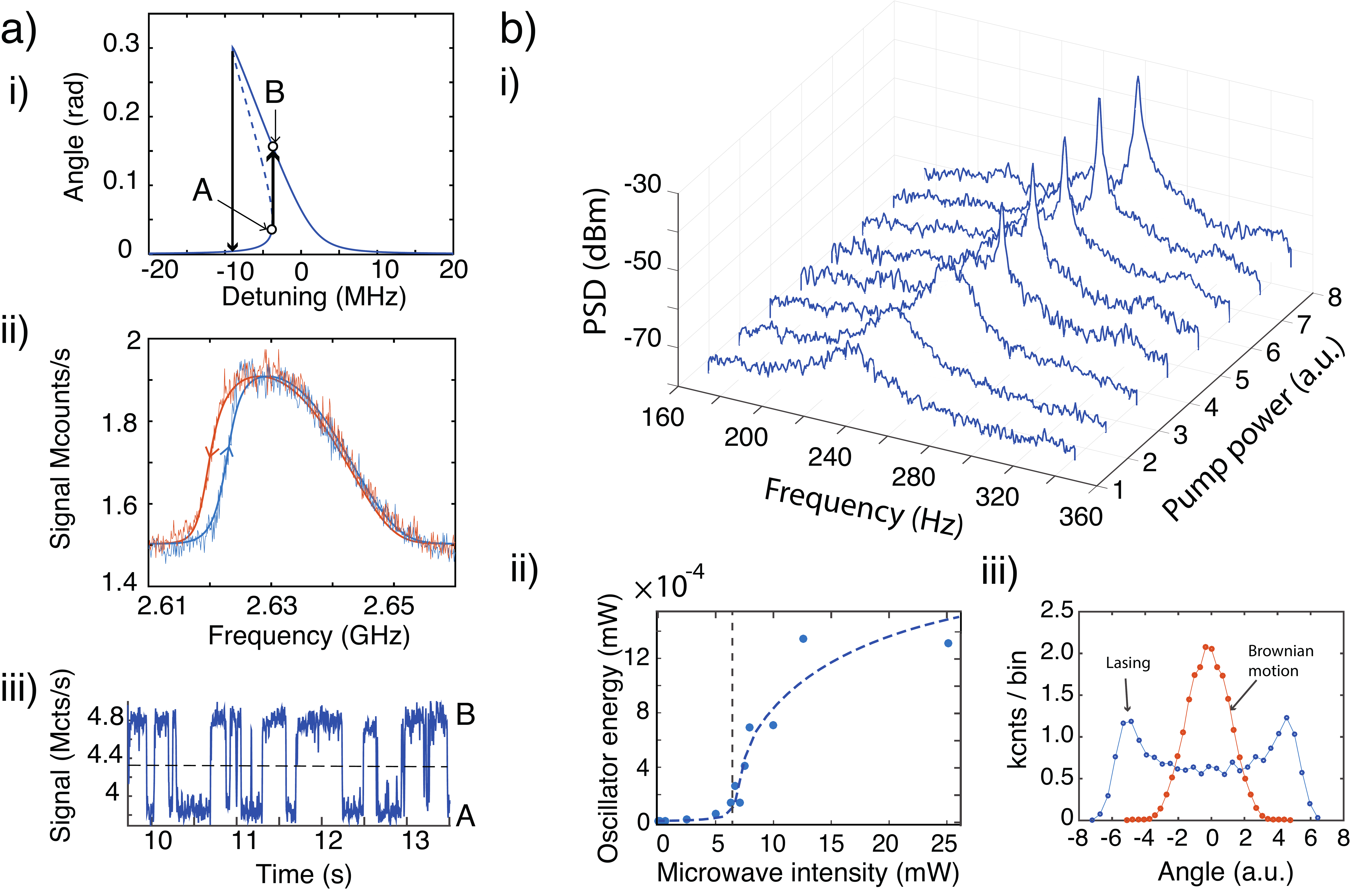}}}
\caption{{\bf Non-linear spin-motion dynamics} a) {\bf Bistability} i) Evolution of the particle angle as a function of the microwave detuning. 
ii) Hysteresis behavior of the particle orientation when the microwave signal is scanned from the red to the blue (blue curve) or from the blue to the red (red curve), as indicated by the arrows. 
iii) Particle orientation as a function of time for a fixed microwave tuned to the red side of the spin transition, showing angular jumps between two stable sites A and B. b) {\bf Phonon lasing} i) Evolution of the power spectrum of the librational motion as a function of the microwave power towards lasing. ii) Oscillator energy as a function of microwave intensity
using microwave powers ranging from -44 up to -16 dBm in steps of 4 dBm. The dashed lines are a fit to the data.  iii) Histogram of the Brownian and lasing angular motions.}
\label{bista}
\end{figure*}

We now make a step into a regime where the spin-mechanical interaction induces non-linear effects on the librational mode. With a stronger spin-torque (see Methods), Fig.~\ref{bista}-a)-i) displays the expected bistable behavior for the angular degree of freedom when the microwave is scanned across the spin resonance. The angle can be found at two metastable positions A or B depending on the history of the angular trajectory (see SI). The hysteresis behavior is indeed observed in the experiment, and shown in Fig.~\ref{bista}-a)-ii). The evolution of the particle orientation over time at a fixed microwave tone is also plotted in Fig.~\ref{bista}-a)-iii). Interestingly, the particle orientation jumps from site A to B in a seemingly unpredictable manner due to random kicks given to the particle. The average population at the angular position B can also be studied as a function of different microwave detunings and increases as the microwave is tuned towards resonance (see extended data and SI).

Let us now set the microwave frequency to the blue side of the spin-resonance in this strong spin-torque regime.
Fig. \ref{bista}-b)-i) shows the power spectral density as a function of the microwave pump power, where a transition from Brownian motion to a self-sustained oscillation is observed. Such a lasing-like action of a mechanical oscillator was observed in the first radiation pressure cooling experiments \cite{Arcizet} with proposed applications in metrology. The spin-mechanical gain that enables such lasing action here is provided by blue detuned microwave excitation which amplifies the angular motion up to a point where losses are compensated by the magnetic gain. 
The oscillator energy as a function of the microwave power is shown in Fig. \ref{bista}-b)-ii). Lasing threshold is observed at 6~mW of microwave excitation. Another signature of mechanical lasing is shown in Fig. \ref{bista}-b)-iii), which displays the probability distribution (PD) of the angular degree of freedom with and without microwave. Under blue detuned microwave excitation, the probability distribution departs from the Gaussian process (red curve) for Brownian motion, and turns to the characteristic PD of a coherent oscillation (blue curve). This effect shows that the librational mode can operate stably, deep in the non-linear regime and highlights further the analogy between the present spin-mechanical platform and opto-mechanical systems. 

Coupling individual spins to the motion of a macroscopic oscillator will have far reaching applications in fundamental science, quantum information and metrology. 
%The present results show that the long-lived spin of NV centers can actuate the temperature of a mechanical oscillator, in strong analogy with laser cooling of trapped atoms. 
The present spin-dependent torque itself may be employed for detecting atomic defects with electrons spins that cannot be efficiently detected through ODMR.
%Such a controllable micro-compass will find implications in torque sensing of trapped particles \cite{tongcangPRL}. 
Further, the approach may also be applied to other torsional nano-mechanical platforms \cite{WuM, Huillery}, which can exploit the long NV spin-lattice relaxation at low temperatures for longer interrogation times and efficient cooling. Last, operating in the sideband resolved regime where $\omega_\phi/2\pi \gg 1/T_2^*$ can be realized after modest improvements to the present set-up. 
We estimated that using a 1~$\mu$m diameter pure diamond grown by chemical vapor deposition (CVD) attached to a 1~$\mu$m diameter ferromagnet, would enable entering the sideband resolved regime for this hybrid structure.
Librational frequencies $\omega_\phi/2\pi$ above 200 kHz have indeed been observed recently \cite{Huillery} and NV centers with $1/T_2^*=50$ kHz electron-spin decoherence rates can readily be obtained in CVD grown microdiamonds enriched in $^{12}$C.
Entering this regime would offer immediate prospects for ground state cooling the diamond libration and for multipartite spin-entanglement and would provide strong impetus to bridge the gap between trapped particles and trapped atoms.\\

\vspace{12pt}

\noindent

{\bf Acknowledgments}
We would like to thank R. Blatt, C. Voisin, Y. Chassagneux, E. Baudin and S. Del\'eglise for discussions. 
\\
\noindent
{\bf Author Contributions}
T. D. and P. H. contributed equally to this work.
T. D, P. H., L. N. and G. H. performed the spin-torque experiments; T. D, P. H., and G. H. analyzed the data and performed the modeling with assistance from L. N., and G. H., T. D and P.H. wrote the manuscript. All authors contributed to the interpretation of the data and commented on the manuscript.

\let\oldaddcontentsline\addcontentsline% Store \addcontentsline
\renewcommand{\addcontentsline}[3]{}% Make \addcontentsline a no-op


%\let\addcontentsline\oldaddcontentsline

%{\bf Online Content} Methods, along with any additional Extended Data display items and Source Data, are available in the online version of the paper. references unique to these sections appear only in the online paper.
%

%\noindent
%{\bf  Author Information} Reprints and permissions information is available at www.nature.com/reprints. The authors declare no competing financial interests. Readers are welcome to comment on the online version of the paper. Publisher's note: Springer Nature remains neutral with regard to jurisdictional claims in published maps and institutional affiliations. Correspondence and requests for materials should be addressed to G. H.(gabriel.hetet@ens.fr).
%
%
%\noindent
%{\bf The authors declare no competing interests}.

%\documentclass[preprintnumbers,amsmath,amssymb,superscriptaddress,twocolumn,showpacs]{revtex4}
\clearpage

\vspace{0.2in}

\begin{widetext}
\begin{center}
 \Large {\textsc{Spin-Cooling of the Motion of a Trapped Diamond} }\\
  \Large {\textsc{\it{Methods}}} 
%{\center \Large {\textsc{  }}\\
\end{center}

\end{widetext}

\noindent{\bf Microdiamond properties} The diamonds that we employ are in the form of a powder of particles that have a diameter of 15 $\mu$m. They are supplied by the company Adamas, which produces diamonds with a concentration of NV centers in the 3-4 ppm range, corresponding to 1.5 to 2$\times 10^9$ NV centers per microdiamond. 
Using the same collection optics as was used in \cite{SI_DelordPRL} we could also estimate the number of optically addressed NV centers to lie in that same range.
This is 4 to 5 orders of magnitude larger than the concentration compared to the experiments reported in \cite{SI_DelordPRL,SI_vacuumESR,SI_delord2016}, where no spin-torque was observed. Under continuous laser excitation at around 10~$\mu$W, the diamond starts to heat up at 0.1 mbar, at similar pressures than in \cite{SI_vacuumESR} which points towards the role played by other impurities than the NV centers in the heating observed in \cite{SI_vacuumESR}.
The pressure we operate at is 1 bar for the spin-torque measurements shown in Figure 2 in the main text, and in the mbar range for the cooling and phonon-lasing experiments.  

\noindent{\bf The Paul trap}
We operate with a Paul trap that is similar to the one used in \cite{SI_DelordPRL} except that the particles are stably trapped at the bottleneck region of the trap, where both the electric field gradient and anisotropy are stronger, yielding higher librational confinement. 
Below 1 mbar, the diamond starts to rotate due to a locking mechanism induced by the Paul trap drive, making it impossible to observe the spin-dependent torque, which relies on very stable libration. 
%The NV spins within the levitating diamond can be manipulated by driving a microwave current through the trap electrode similarly to \cite{SI_vacuumESR}. 

\noindent{\bf NV spin polarization and read-out} Due to an intersystem crossing in the excited state of the NV centers, the electronic ground state $\ket{0}$ is brighter then the ground state $\ket{1}$ upon green laser optical illumination. 
%Even though the magnetic field direction is not aligned with the NV axis, and can thus mix the spin states, with the moderate magnetic field we use the ground state will still be brighter than the excited spin state . 
This provides a means to read out the Zeeman splitting by scanning a microwave tone around the resonance, carrying out Optically Detected Magnetic Resonance (ODMR).
Here, the microwave is applied directly on the trapping electrode, which provides an efficient means to excite the spins. 
The photoluminescence is detected using standard confocal microscopy. We use about 100 $\mu$W of laser light at 532 nm to polarize the NV centers. The laser is focussed via a lens inside the vacuum chamber which has a numerical aperture of 0.5 and a working distance of 8 mm. The focal point of the laser is kept few tens of micrometer away from the microdiamond to mitigate the effect of radiation pressure and to enable laser excitation of the whole diamond \cite{SI_DelordPRL,SI_delord2016}. 
To measure polarization rate to the ground and to the excited magnetic states, we carry out the sequence depicted in Fig. \ref{polar}-a).
The photo-luminescence (PL) rate is measured as a function of the time $\tau$ for both sequences and is plotted in Fig. \ref{polar}-b).
The laser induced polarization rate to the $m_s'=0$ state is 3.3 kHz. The microwave polarization rate $\Gamma_M=\Omega^2 T_2^*$ (see SI) to the magnetic state $m_s'=-1$ is found to be 8 kHz when using -5 dBm of microwave power measured before a 25 dBm amplifying stage.  
An estimation based on both the ODMR width and a Ramsey sequence yields $T_2^*=70$~ns, implying $\Omega/2\pi=60$ kHz.
%Note that the microwave intensity is not exactly the same across the whole spectrum. The gain in the microwave generation chain (that includes SMA cables, a microwave switch, a high power amplifier, analog electronics and then the antenna) is indeed not uniform across the spectrum. This explains the differing heights of the ms=+1 and ms=-1 peaks in the ODMR spectra in Fig.1

The exact degree of spin polarization cannot be estimated precisely without using a full numerical model and the 8 rate equations including mixing by the magnetic field transverse component.  The magnetic field transverse component reduces the polarization time due to mixing of electron spin states both in the ground and excited levels. This enhances the probability of non-radiative crossing to the metastable level and reduces the ODMR contrast by 30$\%$\cite{SI_tetienne_magnetic-field-dependent_2012}.  Overall, this reduces the degree of optical polarization to the $m_s'=0$ spin state to about 60 $\%$.

\noindent{\bf Detection and analysis of the librational modes} The diamond motion is detected by collecting the back reflected green light from the diamond interface, separated from the excitation light using a polarizing beam splitter. The best sensitivity is achieved by taking advantage of the speckle pattern produced by the rough surface of the microdiamond under coherent illumination. At the particle image plane, which is located a few tens of centimeters away from the particle, an image is formed with an additional speckle feature. To detect the diamond motion, we focus a small area of this image onto a single-mode optical fibre and detect the photons transmitted through the fibre with a single-photon avalanche photodiode. The detected signal is then highly sensitive to the particle position and orientation.

\noindent{\bf Angular displacement sensitivity}
For a given levitating particle, we can optimize, in real time, the signal coming from the angular displacement of the particle by selecting the most favorable region of the particle image. To do this, we look at our optical signal while switching a microwave field tuned to one ODMR transition at a frequency of 1 Hz. Alignment is done by maximizing the change in the coupled light intensity as the diamond jumps between two angular positions. The linearity of the coupled light with the rotating angles can finally be assessed by looking at higher order of the harmonic motion once the librations frequencies are identified.

While being a sensitive measurement of the angular displacement, our optical signal is not an absolute measurement. 
The spin-torque vector $\vec{\Gamma_s}$ is orthogonal to the plane defined by the magnetic field and the NV axis (it tends to align the NV axis to the B field). However, because the angular confinement is not isotropic, the particle rotation angle is not necessarily collinear to the spin-torque. Determining the exact three-dimensional rotational dynamics of the particle would necessitate knowledge about the orientation of the NV axes with respect to the principal axes of the angular motion.

Using NV magnetometry, the mechanically detected spin resonance can nonetheless be used to relate the optical signal change to the angular displacement of the particle.  A set of three mechanically detected spin resonances corresponding to three different microwave powers are shown in Fig. \ref{calib_angle}-a) of the extended data, under a magnetic field of 144 G.  The minima of each curve fall on the dashed line.  The lower panel of Fig. \ref{calib_angle}-a) is a theoretical curve where the angle between the NV axis and the magnetic field direction is plotted as a function of the frequency of the NV spin transition. This curve is obtained by diagonalizing the NV spin Hamiltonian in the presence of a magnetic field of 144 G. 
Since the maximum magnetization of the NV spins is obtained when the microwave field is resonant with the spin transition, one can relate the maximal change in the optical signal ($\Delta S$) to the variation of the angle between the NV axis and the magnetic field direction ($\Delta \theta_{NV}$). Doing so, we obtain here a resolution of 43~mrad/Mcounts/s. Fig. \ref{calib_angle}-b) shows a time trace of the optical signal upon brownian motion of the particle. From the standard deviation of this signal and the above calibration, we obtain an angular displacement sensitivity of 0.3~mrad/$\sqrt{\rm Hz}$. 

These numbers are however only upper bounds for our resolution and sensitivity.
To explain why this is the case, Fig. \ref{calib_angle}-c) shows a sketch of the angular motion of the diamond after magnetizing one class of NV spins. For simplicity, we consider rotation about two axes here. In a referential frame with axes given by the principal librational mode directions, we can parametrize the orientation of the NV axis in a subspace defined by the two angular coordinates $\theta_x$ and $\theta_y$. The orientation without magnetization ($M_z=0$) is given by the trap and particle geometry and labelled O. The point B in this space is the direction of the magnetic field.
Upon magnetization, a torque is applied to the particle such that the orientation follows the OB trajectory over time. However due to different confinement of the librational modes $\omega_x$ and $ \omega_y$ along the $x$ and $y$ axes, the angular motion takes place along a different trajectory. 

In our experiments, the orientation of the magnetic field and NV axes relative to the principal axis of the libration is unknown. This prevents us from fully calibrating our detected angular motion. Nevertheless, provided that the detection is optimized to the librational mode having the highest confinement, we can ensure that the detected angular displacement $\theta_d$ is smaller than the angular displacement $\theta_{NV}$ sensed by the NV spins.
This can be seen in Fig. \ref{calib_angle}-c), where we note NV the equilibrium position when $\omega_x>\omega_y$. Our calibration method thus gives an upper bound to the obtained resolution and sensitivity. 
The optimization of the detection is performed by monitoring the power spectrum and tuning the speckle angle at the entrance of the fiber to maximize the power spectrum of the mode with the largest frequency.
Fig. \ref{calib_angle}-d) shows the power spectrum of the brownian motion for two detection alignments. In red, all three librational modes, indicated with the black arrows, are clearly visible. In blue, the detection is tuned to be mainly sensitive to the mode with the highest confinement frequency. The latter detection tuning is used for the data shown in a) and b).

The sensitivity could be improved by collecting all the speckle pattern using a SPAD camera rather than just a fraction of it as we do now. Using a shorter laser wavelength would also straightforwardly improve the sensitivity. Another technical limitation comes from the trapped diamond motion in other modes than the libration mode of interest which adds noise to the angular displacement signal. In this regard, active stabilization of the center of mass will greatly increase the sensitivity.
   
\noindent{\bf Power Spectral density} Using the above described detection method, motional frequencies can be observed by sending the detected signal to a spectrum analyzer. Under vacuum (1 mbar), the power spectral exhibits narrow peaks at the trapping frequencies of the motional modes which are driven by the Brownian motion (see Fig 1-b) in the main text). For each librational mode, the power spectrum is fitted by the formula obtained in the SI :
\begin{eqnarray}
S_\phi(\omega)&=&\frac{2\gamma kT}{I ((\omega_\phi^2-\omega^2)^2 +\gamma^2\omega^2)}.
\end{eqnarray}

The librational modes can in fact be unambiguously identified (and separated from the center of mass modes) using the NV centers induced torque. By switching on and off a microwave field tuned to one spin resonance at the same period as that of one diamond libration, one performs parametric excitation of that librational mode. In our experiments, a sequence of five microwave pulses is enough to displace the angle above the Brownian thermal noise. Following a parametric excitation sequence, the diamond orientation "ring-down", or decay, is observed. A typical decay curve is shown in Fig. 4-a). We typically find librational frequencies in the 100~Hz to 1~kHz range.

\noindent{\bf Parameters used for the spin-dependent torque measurements} The ODMR and spin-mechanical measurement scans shown in Fig 2-a) of the main text are taken under atmospheric pressure. The green laser power was 330~$\mu$W and the microwave power set to 0~dBm. The magnetic field is around 95 G. 
For the mechanically detected spin resonance of trace ii), the microwave detuning is scanned in 2~MHz steps with a duration of 10 ms per points. During those 10ms, the diamond orientation has enough time to reach its equilibrium position and the spin torque effect can be observed. The average count-rate is 2.3$\times10^6$~s$^{-1}$ for a total averaging time of 10 minutes.
For the ODMR trace iii), the microwave detuning is scanned by step of 2 MHz with a duration of 1ms per points. For each point, the microwave field is switched off for the first 0.5ms and switched on for the last 0.5 ms during which the signal is acquired.
This prevents mechanical effects from altering the detected NV centers photo-luminescence (PL) signal. The PL count rate is 
5$\times 10^6 $~s$^{-1}$ and the total averaging time for this measurement was 3h.

\noindent{\bf Estimation of the temperature} Measurement of the temperature associated with the librational modes can only be an estimate. Obtaining a precise temperature value would require knowledge about the moment of inertia of the particle, which is prone to strong systematic errors. The standard method is to vary the pressure \cite{SI_Hebestreit} while observing the power spectrum : over the pressure range where its area is constant (to satisfy Liouville's theorem under adiabatic transformation) the librational mode temperature is known to be 300 K as it is thermalized with the gas temperature. In our case, pressure variations slightly change the orientation and position of the trapped particle and, incidentally, the sensitivity to angular motion. This prevents such method from being used. However several measurements support the fact that the external degrees of freedom of the particle are thermalized at 300 K when operating in the mbar pressure range. First, we measured the particle internal temperature with our typical laser powers {\it via} NV thermometry \cite{SI_vacuumESR} and found it to be close to 300 K. This insures that no heating of the libration modes comes from the heating of the gas surrounding the particle \cite{SI_millen2014nanoscale}. We observed that heating of the particle by the laser starts below 0.1 mbars, similar to what was measured using diamonds that were doped with a three orders of magnitude smaller NV concentration. 
Several sources of noise could also heat up the particle, such as the laser induced torque \cite{SI_delord2016} or charge fluctuations. Heating by the former can be excluded as no noticeable changes of the power spectrum shape occur when laser power is increased up to 1~mW.

\noindent{\bf Parameters and calibration used for the cooling measurements} The power spectrum of the detected librational modes depends strongly on the particle angle. For the same motional amplitude, a change in the particle angle potentially implies a different speckle pattern, which in turn changes the power spectrum sensitivity.
Since the particle angle changes with the microwave detuning and power, different power spectra cannot be directly compared when the parameters are changed. The three traces i), ii) and iii) in Fig. 2-b) in the main text have thus been obtained at the same particle angle to enable quantitative comparison between their areas. 
Operating at the same particle angle was ensured by performing a resonant spin-mechanical detection at different microwave amplitudes and choosing pairs of microwave frequencies and powers that correspond to the same count rates. As shown in the data in Fig. 4-b), we chose microwave detunings corresponding to the points 1-2-3 for the two traces i) and ii) taken at microwave powers of -20~dBm and -10~dBm respectively. The frequencies, which are 2.617, 2.623 and 2.634~GHz, respectively to the blue, resonant and red side of the spin resonance all correspond to the same angle under these power conditions. 
A fit to the experimental curves in Fig. 2-b) was obtained using the formula 
\begin{eqnarray}
S_\phi(\omega)&=&\frac{2\gamma_{\rm eff} kT}{I ((\omega_{\rm eff}^2-\omega^2)^2 +\gamma_{\rm eff}^2\omega^2)}.
\end{eqnarray}
The dependence of the damping $\gamma_{\rm eff}$ and frequency shift $\omega_{\rm eff}$ with the microwave detuning shown in Fig.~2-c) was obtained using parametric excitation of the librational mode at 480~Hz. The microwave power is $-10$dBm and for this measurement, the above mentioned calibration issue (change in the sensitivity when the microwave detuning is varied) is not relevant.
In order to extract the damping and shifts, the resulting ring-down was fitted by the formula
\begin{eqnarray}
S(t)&=& A_1\text{sin}(\omega_{\rm eff}t+\phi)\text{exp}(-\gamma_{\rm eff}t /2)\\
&+& A_2\text{sin}(\omega_2t+\phi_2)\text{exp}(-\gamma_2 t / 2) + A_0, \nonumber
\end{eqnarray}
where the second exponentially damped sinus takes into account the slightly excited librational mode at 590~Hz.
Three of these ring-down traces are shown in the Fig. 4-c). For each curve, the averaging time is around 100~s. An estimation of the temperature  relies on comparing the damping with and without spin-cooling, using the relation $T_{\rm eff}=T\frac{\gamma}{\gamma_{\rm eff}}$ \cite{SI_Arcizet2}.

\noindent{\bf Modeling of the experiment}  For the most part of the paper, we model the experiment numerically using Monte-Carlo simulations that include the full three-level structure of the NV spin 1 system in the ground state (see Supplementary Material).
For the spin-cooling and spin-spring effects shown in Fig. 2.c), the number of NV centers, polarization rate, Rabi frequencies, and angle between the NV centers and the main axis of the diamond are left as free parameters.

\noindent{\bf Experimental analysis of bistability and phonon-lasing} The curve in Fig. 3-a)-i) shows the angular evolution of the particle as a function of the microwave detuning obtained using similar parameters than in the linear regime, but using a microwave and laser powers that fitted best the data in Fig.3-a)-ii) and a lower trapping frequency ($\omega_\phi/2\pi=240$~Hz). Fig.3-a)-iii) shows the evolution of the librational mode angle as a function of time in this regime. Several of such curves were obtained for different microwave detunings, and are shown in the extended data. A Monte-Carlo simulation was also performed using our experimental parameters, with a microwave tuned to the red side, and shows similar jumps between the two stable points A and B.
%The sensitivity of the measurements for different microwave powers has no impact in this analysis and the fit to the data was done using the model presented in the SI. 
The data shown in Fig. 3-b)-i) of the main text show the evolution of power spectra for different microwave signal powers, when it is tuned to the blue of the ODMR transition. 
The onset of instability is seen at approximately 0 dBm. For a quantitative estimate of the threshold, we compute the area below the librational peak as a function of microwave detuning. This is shown in Fig. 3-b)-ii). Note that here the sensitivity of the power spectra to the angle may induce some systematic errors. For these measurements, we fitted the data by the numerical model, and found good agreement with the numerical analysis, but a quantitative comparison with the experiment is difficult due to the above mentioned angle dependent sensitivity. 
%The estimation of the threshold and the deviation of the lasing behavior from Gaussian statistics (shown in Fig. 3-b)-iii)) depends only weakly on the microwave power dependent change in the sensitivity however. 

%\begin{multicols}{1}. 
%\maketitle 

\clearpage

\vspace{0.2in}

\begin{widetext}
\begin{center}
  \Large {\textsc{\it{Extended data}}} 
%{\center \Large {\textsc{  }}\\
\end{center}

\begin{figure}[ht!]
\centerline{\scalebox{0.135}{\includegraphics{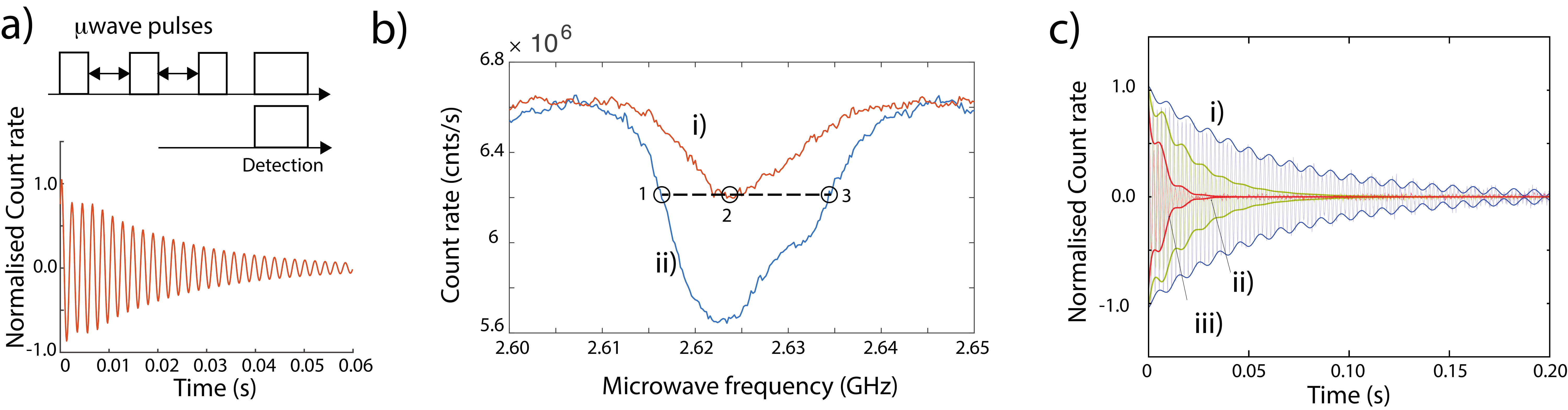}}}
\caption{a) Detection principle of the librational modes using microwave parametric excitation (top). 
Reflected light field amplitude as a function of time for one low librational mode frequency (170 Hz) at 1 mbar (below).
b) Spin-mechanical resonance for two different microwave powers i) -20 dBm and ii) -10 dBm.
c) Reflected light field amplitude as a function of time, for three microwave frequencies  $2.617, 2.623$ and $2.634$ GHz. Plain line shows a very good fit to the data (see Methods) using two sinus at the frequency of the parametrically excited librational mode (480 Hz) and of the closest librational mode (590 Hz) with an amplitude that is 30 times smaller. 
}\label{angularsplitting}
\end{figure}

  \begin{figure}[ht!]
\centerline{\scalebox{0.3}{\includegraphics{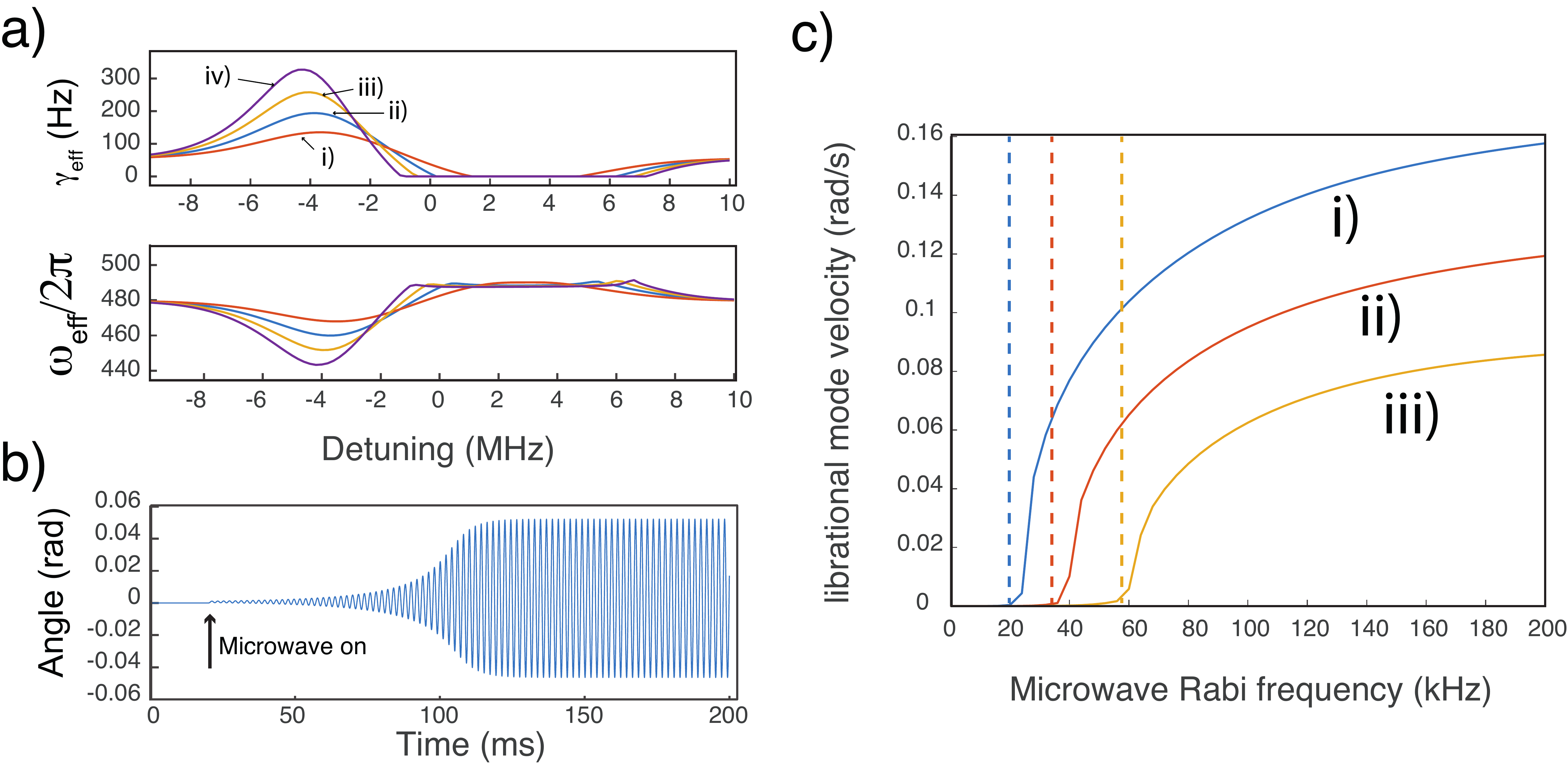}}}
\caption{Numerical simulations in the non-linear regime. a) Effective damping and spring constants as a function of the microwave detuning deduced by a numerical fit to the ring down curves (see Methods). Trace i), ii), iii), and iv) correspond to the increasing microwave Rabi frequencies $\Omega/2\pi=50,150, 200, 250$~kHz respectively. b) Evolution of the angle as a function of time after turning on a microwave to the blue of the spin resonance at a time $t=20$~ms, showing amplification and finally lasing at $t=120$~ms. c) Librational mode velocity as a function of microwave Rabi frequency, for three different values of the spin-lattice relaxation rate $1/T_1=1, 2$ and $3$ kHz for trace i)-ii) and iii) respectively.}\label{unstable}
\end{figure}

\begin{figure}[ht!]
\centerline{\scalebox{0.65}{\includegraphics{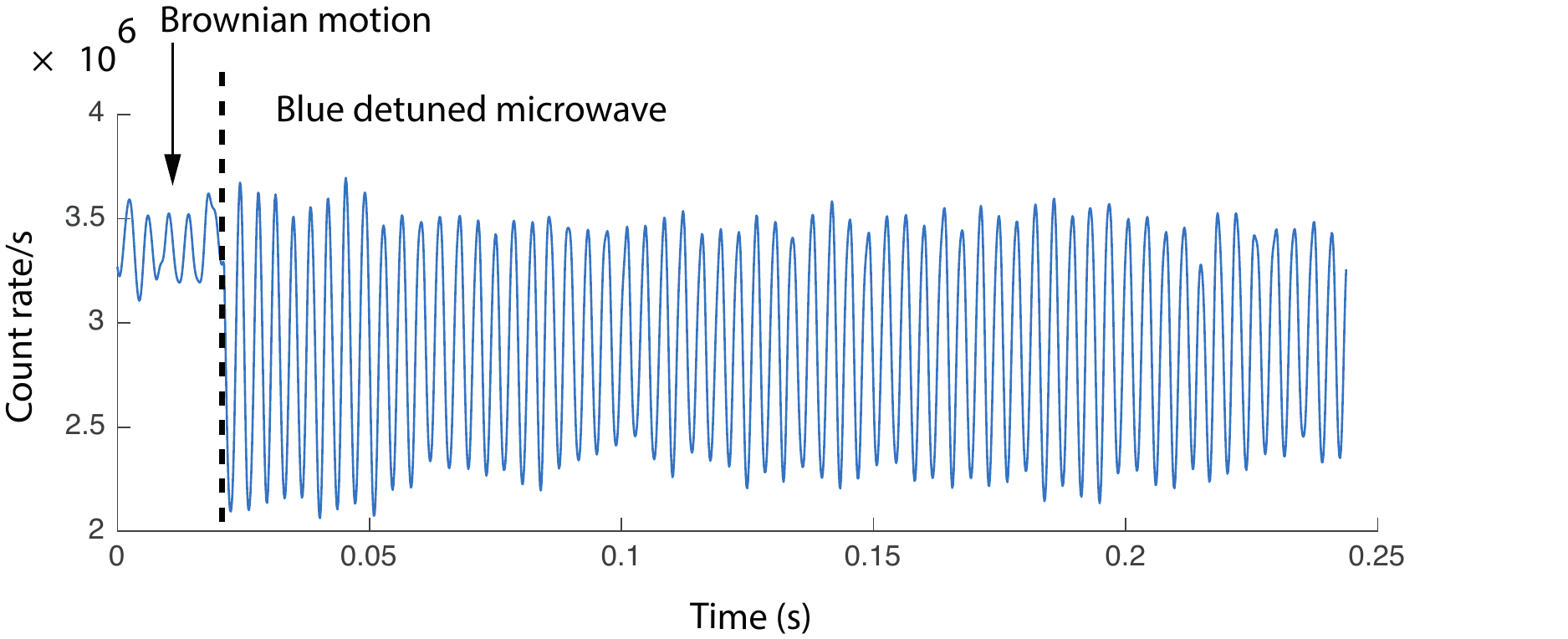}}}
\caption{Particle angle as a function of time upon sudden switching on of a microwave signal at a time t=0.02 s on the blue of the spin transition. The microwave power is above threshold so lasing can be observed. This curve was used to plot the histogram of Fig. 3-c) in the main text.}\label{hisogramme}
\end{figure}

\begin{figure}[ht!]
\centerline{\scalebox{0.5}{\includegraphics{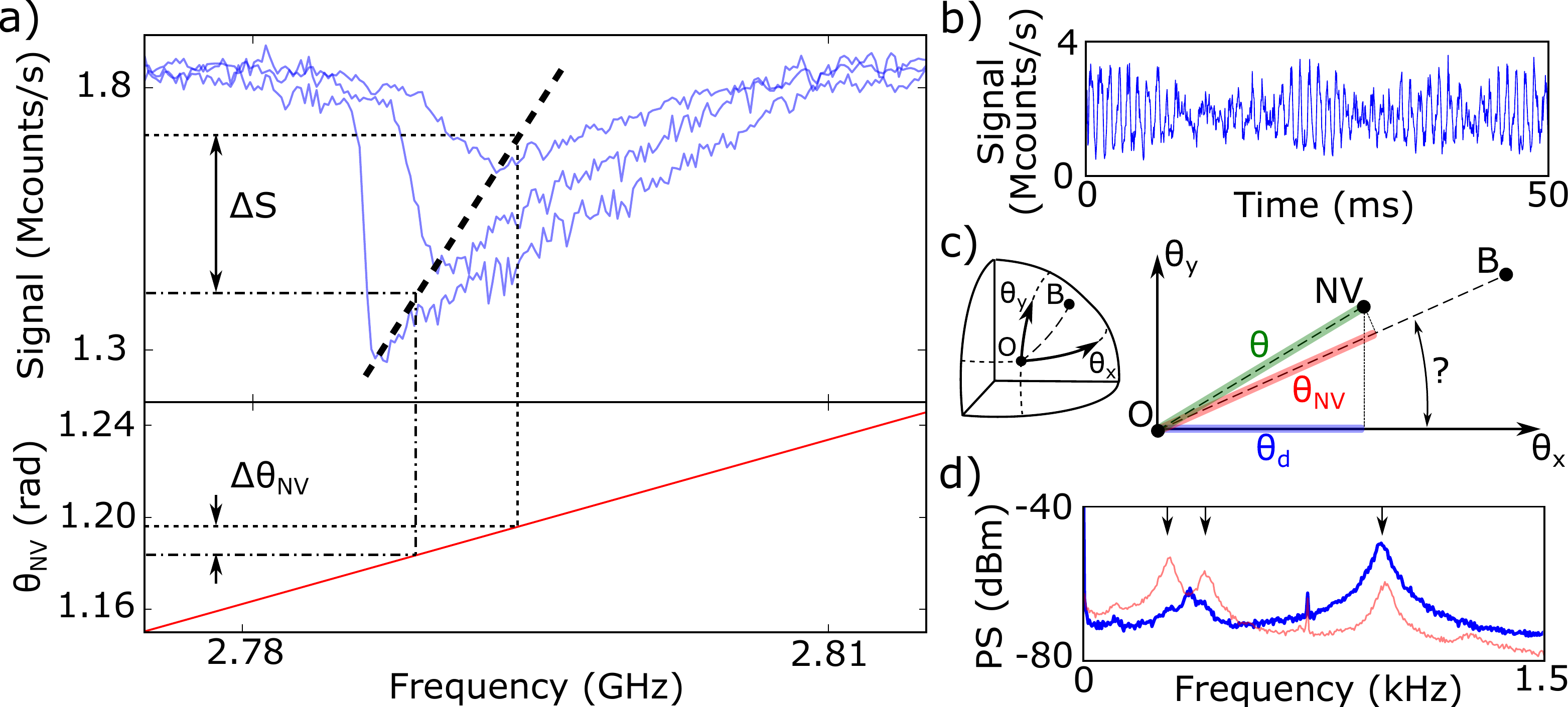}}}
\caption{a) Upper panel: mechanically detected spin resonance for 3 different microwave powers. 
The dashed line is the locus of the signal minima. 
Lower panel: angle between the NV axis and  magnetic field direction versus NV spin transition frequency. 
b) Optical signal as a function of time. c) Left panel : Sketch depicting the angular motion of the diamond upon magnetization of one class of NV spins using only two angles $\theta_x$ and $\theta_y$ for simplicity. B is the orientation of the magnetic field. 
Right panel : angular trajectory represented in the ($\theta_x$, $\theta_y)$ space. 
We note O the particle orientation without NV magnetization ($M_z=0$). The red/green line are the trajectories in the isotropic/anisotropic case respectively (see Methods).
$\theta_d$ is the detected angle.
d) Power spectrum of the Brownian motion for two different speckle alignments taken with a resolution bandwidth of 1 Hz. The red curves show all three librational modes. 
In blue, the detection is tuned to be mainly sensitive to the mode with the highest confinement frequency. The latter detection tuning is used for the data shown in a) and b).}\label{calib_angle}
\end{figure}

  \begin{figure}[ht!]
\centerline{\scalebox{0.4}{\includegraphics{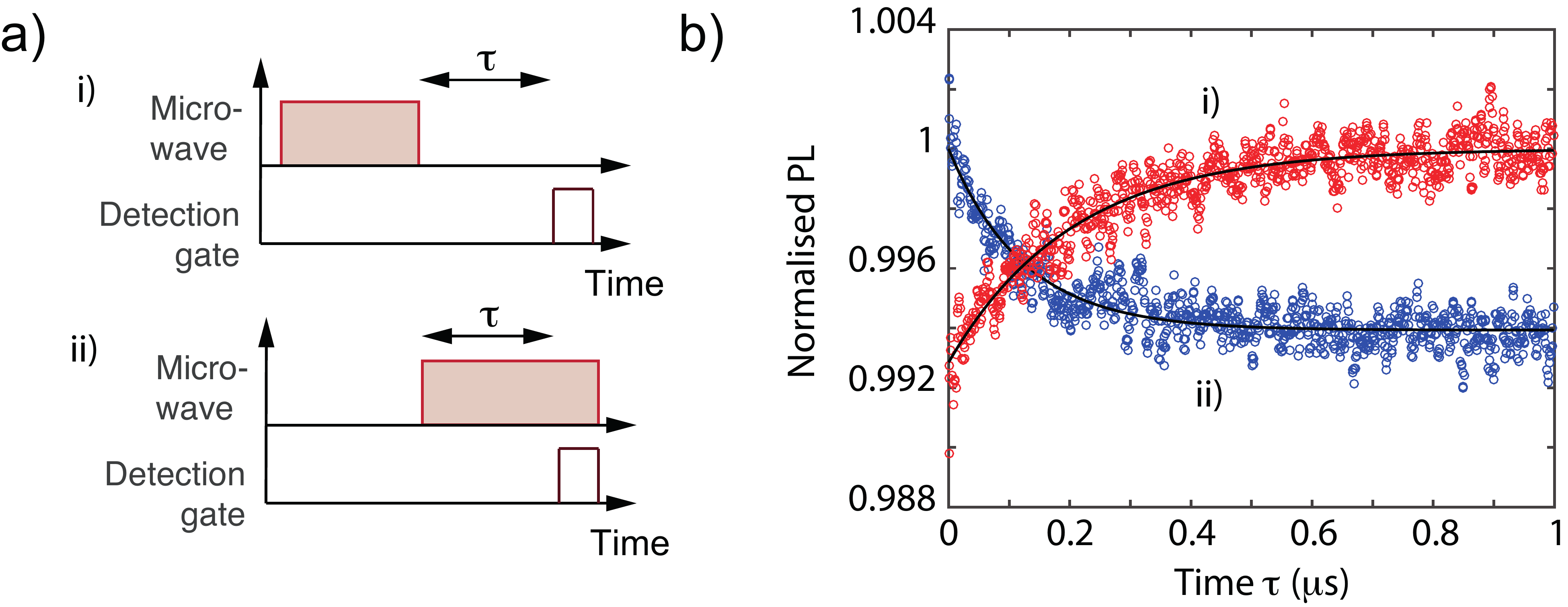}}}
\caption{a) Sequences employed to measure the laser induced polarization rate to the ground state (trace i)) and  to measure the polarization rate in the magnetic state $m_s'=-1$ {\it via} the microwave. The laser is kept on at all times for both sequences.
b) Trace i) shows the increase in the PL rate after turning off the microwave signal. An exponential fit to the data gives a laser polarization rate of 300 $\mu$s. Trace ii) shows the decrease in the PL rate after turning on the microwave signal. The polarization rate to the magnetic state is 124 $\mu$s.
}\label{polar}
\end{figure}

\begin{figure}[ht!]
\centerline{\scalebox{0.45}{\includegraphics{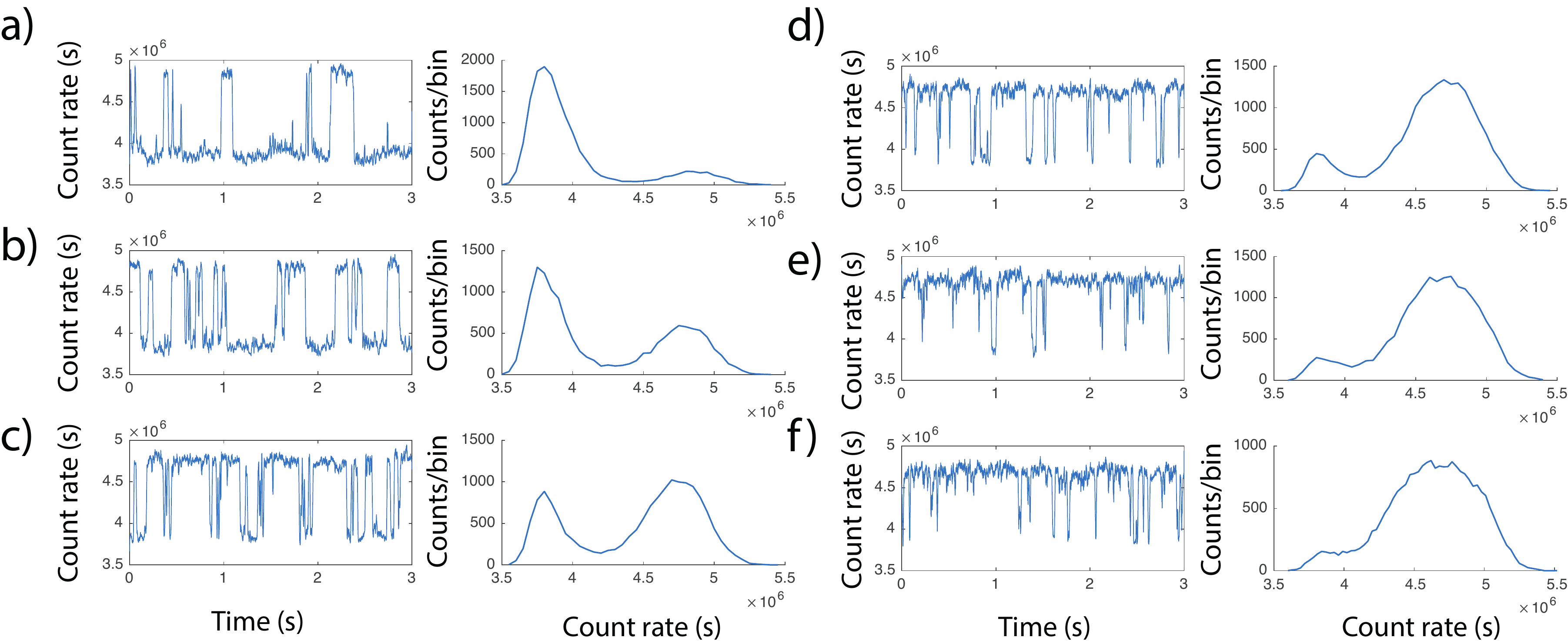}}}
\caption{Experimental observation of the particle angle as a function of time for different microwave frequencies for microwave frequencies : 2.8000, 2.8010, 2.8020, 2.803, 2.8035, 2.804 GHz, from trace a) to f), tuned to the red of the ESR transition. A histogram is shown on the right of each time-trace, showing the number of counts at each count-rate bin.}\label{bistamicro}
\end{figure}

\end{widetext}

\clearpage

\begin{widetext}

\vspace{0.2in}

\begin{center}
 \Large {\textsc{Spin-Cooling of the Motion of a Trapped Diamond} }\\
  \Large {\textsc{\it{Supplementary Material - Spin-cooling and phonon-lasing theory}}} 
%{\center \Large {\textsc{  }}\\
\end{center}

%\begin{multicols}{1}. 
%\maketitle 

%\end{multicols}. 

 \subsection{The mechanical oscillator}

The considered mechanical oscillator is the librational mode of a microdiamond levitating in a Paul trap. 
It was shown in \cite{SI_delord2016,SI_DelordSt} that the Paul trap induces a confinement of the angle due both the trap and diamond particle asymmetry. In the manuscript, we directly observe the three librational modes under vacuum using the speckle detection detailed in the Method section. 
We fit the power spectral density of the librational modes in Fig. 1-b) of the main text using the equations that are derived from the following linear response theory.

In the following we consider the simple case of a single excited angle, assuming that the other angles are not coupled to it.
The librational motion of the diamond in the Paul trap is ruled by the equation of motion for the angle between the main axis of the diamond and the main axis of the trap \cite{SI_DelordSt}. Noting this angle $\phi$, we get 
\begin{eqnarray}I \ddot \phi&=& -I \omega_{\phi}^2 \phi -I\gamma \dot \phi+\Gamma_T(t)
\end{eqnarray}\label{eqnmotion}
where $I$ is the moment of inertia, $\omega_{\phi}$ is the angular frequency, $\gamma$ is the damping rate due to collisions with the background gas and $\Gamma_T(t)$ is the associated Langevin torque.

Fourier Transforming this equation yields 
$$\phi(\omega)=\chi(\omega)\Gamma_T(\omega)$$
where
$$\chi(\omega)=\frac{1}{I(\omega_{\phi}^2-\omega^2+i\omega \gamma)}.$$
The Langevin torque obeys the relation

$$\langle \Gamma_T(\omega)\Gamma_T(\omega')\rangle=2\pi \delta(\omega+\omega')S_T(\omega)$$
where 
$$S_T(\omega)=-\frac{2kT}{\omega}{\rm Im}\Big[\frac{1}{\chi(\omega)}\Big]$$ when the number of phonon excitations is larger than $1$.
One finds
 $$S_T(\omega)=2kT\gamma I.$$
The librational spectrum is then found to be 
\begin{eqnarray}
S_\phi(\omega)&=&|\chi(\omega)|^2 S_T(\omega)\\ \nonumber
&=&\frac{2\gamma kT}{I ((\omega_{\phi}^2-\omega^2)^2 +\gamma^2\omega^2)}\nonumber.
\end{eqnarray}
This formula describes very well the observed librational motion shown in Fig. 1 b) of the main text.
Integrating this expression over $\omega$, we obtain
$$\frac{1}{2} I \omega_{\phi}^2 \langle \phi^2\rangle= \frac{1}{2} kT $$
where $$\langle \phi^2\rangle=\int S_\phi(\omega) d\omega,$$
in agreement with the equipartition theorem. 
In principle, the area below the Lorentzian curves observed in Fig. 1 b) gives us direct access to the temperature (see Methods, for a discussion on temperature calibration).

\subsection{Rotating a levitating diamond using the spin of nitrogen-vacancy centers}

We now discuss the theory behind the spin-induced torque from the nitrogen-vacancy centers.
In the secular approximation of the Paul trap potential, the Hamiltonian describing the angular motion takes the form
\begin{equation}\begin{aligned}
H_{\rm meca}=\frac{1}{2}I\omega_{\phi}^2\phi^2 + \frac{L^2}{2I}.
\end{aligned} \end{equation}
In our experiment, the angular position of the particle in the trap is shifted due to a torque induced by spins of NV centers in the presence of a magnetic field, which will add a magnetic potential energy.

%To simplify the approach, we consider a 2D model where the spin and magnetic field remain in the same plane perpendicular to the libration mode rotation axis. The internal spin operators $(\hat{S_x} ,\hat{S_z})$ lie on a 2D plane. 

The NV centers is a spin triplet system due to the presence of two paired electrons. The singlet state lies at much higher energies and only contributes to the triplet state polarization via the green laser. The dipolar interaction between the two electron spins lifts the degeneracy between the $|m_s'=0\rangle$ and  $|m_s'=\pm 1\rangle$ states, which sets a preferential quantization axis that is not given by the external magnetic field as it would normally be for free electrons \cite{SI_Doherty_2011}. The degeneracy is already lifted by $D=2.87$ GHz at room temperature in the absence of magnetic field.
This property is at the core of the proposals put forward in \cite{SI_DelordSt, SI_Ma} to let the NV centers act on the angle of a levitating diamond.

Here, we consider a large ensemble of NV centers (about $10^9$ NV centers), which are all along one of the 4 $<111>$ crystalline axes. Under an external magnetic field, the orientation dependent Zeeman effect lifts the degeneracy between these NV spin energies. Therefore when the microwave frequency is close to the resonance corresponding to only one axis, only those spins that are along this axis are in a magnetic state.

To estimate the maximal torque exerted by $N$ magnetized NV centers onto the diamond, one must search for the steepest variation of the Zeeman splitting with the NV angle with respect to the B field.
Let us define the average spin operators as 
$$ \hat S_\alpha=\frac{1}{N}\sum_{i=1}^N \sigma_\alpha^i,$$
where $\alpha$ denotes the three spin directions $x$, $y$ and $z$ and $\sigma_\alpha^i$ are the Pauli matrices of the NV number $i$ corresponding to the direction $\alpha$.

The hamiltonian for the NV ensemble reads 
\begin{equation}\hat{H}_{\rm NV}=\hbar N D \hat{S}_z^2+ \hbar \gamma_e N \bf B  \cdot \bf\hat S
\end{equation}
where $\gamma_e$ is the electron gyromagnetic factor, $\hbar$ the reduced Planck constant and $\bf B$ the external magnetic field. The first term describes the aforementioned spin-spin interaction which is the dominant energy contribution and the second term, the Zeeman energy.

The greatest angular variation of the three eigenenergies with angle can be found between the ground and lowest excited state at approximately $\phi_{\rm opt}=\pi/4$ under moderate magnetic fields (less than 100 G)\cite{SI_Ma}.
The derivative of the energy as a function of the angle can also be estimated close to this angle and can be found to be around $\gamma_e B$.

We can then write the Hamiltonian in the new eigenstate basis, and consider the NV electronic spins as being a simple two-level system in the $\{ |m_s'=0\rangle, |m_s'=-1\rangle \}$ basis. 
Assuming that the spins are promoted to the excited $|m_s'=-1\rangle$ spin state by a resonant microwave, and linearizing the energy dependance with respect to $\phi$ at $\pi/4$, the total energy is changed to 
$$E=\frac{1}{2}I\omega_{\phi}^2\phi^2+\frac{L^2}{2I}-\hbar N \gamma_e B \phi+{\rm cte}$$
The center of the angular potential is thus shifted to
$$\phi_0=\frac{\hbar N {\gamma_e B}}{I \omega_\phi^2}.$$ 
%It is this quantity that was measured by scanning the microwave tone, as shown in the Fig.2-a) of the main text. 
The number of NV centers that are magnetized during a typical microwave scan is at most one fourth of the total number of NV centers due to the 4 involved NV orientations. Including the detrimental influence of the transverse component of the magnetic field on the spin polarization, only $50\%$ of the population is transferred to the $|0\rangle$ state by the laser before being excited by the microwave, so in total, only about $10^8$ NV centers are in a magnetized state.
Taking a $15\mu$m diameter diamond, using typical librational mode frequency of $\omega_\phi / 2\pi = 500$ Hz and an external magnetic field of 100 G, we obtain an angle shift of around $\phi_0=5$~mrad. 

Let us note that the spin-induced angular displacement will significantly increase when decreasing the particle size. Indeed, the moment of inertia scales like the fifth power of the particle diameter. Magnetizing a single spin inside a 500 nm diameter diamond with similar trapping conditions than the presented experiments would lead to an angular displacement of 0.5~mrad.  Our sensitive detection technique (via the speckle) imposes the minimum size of the employed particles to be the illumination wavelength, but detecting such angular displacement within the spin lifetime does not seems out of reach for future experiments.

%In practice, this angular shift can be measured using another NV orientation.
%For such a read-out, using the correspondence $\delta \nu=\gamma_e B \delta \phi$ (if the read orientation is close in frequency to the pumped one) we get 
%$$\delta\nu=\frac{\hbar N {(\gamma_e B)}^2}{I \omega_\theta^2}$$
%The experiment could be done in CW by comparing the angular shift when the microwave pump is tuned in and out of the transition. %In the pulsed regime it would take too much time ($2\pi/\omega_\phi\approx$ 5 ms) for the particle to turn after changing the spin state compared to the coherence time. 

%Including the imperfect polarization $p_0\approx 80 \%$ in the ground state. 
%The end formula for the shift on the probed transition is 
%$$\delta\nu=\frac{\hbar {N p_0(\gamma_e B)}^2}{I \omega_\theta^2}.$$
%
%We need a particle that is larger than 2 $\mu$m in diameter, for stable angular detection, easy injection of single particles, insensitivity to external forces. We also need as much NVs as possible on a single monocrystal, be it attached to 
%another particle or not. 
%

%Close to $\pi/4$, the spin torque is thus $\Gamma_{\rm mag}=\hbar \gamma_e B S_z$.
The above analysis however neglects an important effect that is at the core of the spin-mechanical coupling, that is the back-action of the oscillator onto the spin state. 
The root of this effect is that the magnetic state population also depends on $\phi$. The reason for this is that when the diamond rotates, the B field projection onto the NV axis changes. The microwave will then be brought in or out of resonance, which will in turn change the total magnetization. 
This mechanism is truly analogous to the radiation pressure back-action force onto moveable mirrors in a high-finesse optical cavity. 
There, when the light enters the cavity, it can push the mirror and displace it enough so that the cavity resonance condition is changed.
This change in the resonance condition can in turn 
reduce or enhance the intra-cavity light intensity \cite{SI_aspelmeyer}, which can increase again, or reduce, the radiation pressure force. 
The analogy with the present system is very strong since the interpretation of our experiment can be carried our by essentially replacing the NV spin degree of freedom with the cavity field and the mirror position by the particle angle. 

Let us analyse the mechanisms at play including such, potentially delayed, interaction between the spin and librational degree of freedom.

\subsection{The linear regime : Spin-spring effect and spin-cooling}

The above mentioned spin-mechanical coupling mechanism was discussed in \cite{SI_DelordSt, SI_Ge}.

The analysis of \cite{SI_DelordSt} was carried out for diamonds in a Paul trap in the sideband resolved regime, where the motional frequency is larger than the spin decoherence rate. 
Reducing the mean phonon number in the librational mode can be done in a very similar way than with single ions/atoms by tuning the microwave frequency close to the motional red sideband transition. Fast cooling can be realized by using a green laser that polarises the magnetized spins back to the ground state. In a regime where the spin-excitation rate on the red-sideband is faster than any heating mechanism of the angular degree of freedom (typically collisions with the background gas), such a combination of microwave and laser excitation can lead to ground state cooling of the librational mode. 

In our diamonds, many paramagnetic impurities are present, which induces inhomogeneous broadening of the NV spin resonance transitions by about 7 MHz, making it difficult to reach the sideband resolved regime. Another difficulty is that the particles that we trap have a large moment of inertia (in the $10^{-22} {\rm kg/m^2}$ range).
Technical limitations on the trap parameters (size and voltage) then impose an upper limit to the librational frequency.  

Using a large particle however means that we can use many NV centers so the spin-dependent torque can be very large. Further, although we are far from the sideband resolved regime, since $T_2^*=50$ns and the librational mode period is in the millisecond range, significant cooling of the oscillation can be obtained because of a significant delay between the magnetization of the NV centers (measurements of the magnetization rate are shown in the extended data) and the librational motion. Such spin-cooling mechanism is sketched in Fig. 2-b). 
We now quantify the cooling efficiency using a numerical model.

\subsubsection{Numerical model}

To fit the experimental data Fig.2-c), we study the dynamics of the coupled angle and spin degrees of freedom using Bloch equations.
The green laser induces decay from the $|m_s=\pm 1\rangle$ levels to the ground state $|m_s= 0\rangle$ at a rate $\gamma_{\rm las}$ of about tens of kHz. We note $\Omega/2\pi$, the Rabi frequency of the microwave signal and $\Delta/2\pi$ the frequency detuning with respect to the $|m_s=0\rangle$ to the $|m_s= -1\rangle$ transition at equilibrium.

The dynamics of the paramagnetic spin bath ($\mu$s) is faster than the timescales of the motion and spin-spring shifts so we can treat the paramagnetic impurities as a general Markovian reservoir that limits the spin coherence time $T_2^*$ to 50 ns \cite{SI_DelordPRL}, much below the $T_1$ time of the spin populations ($\approx ms$ in our experiment). 
We write $S$ the mean expectation value of the Pauli operator $\hat \sigma^-_i$ of the individual spins on the subspace \{$|0\rangle$,$|-1\rangle$\}. We also write $\hat S_z^\beta=|\beta\rangle \langle \beta|$, the population in the three spin states $|\beta\rangle$ 
and $S_z^\beta$ its expectation value.
 %$S_z^j=\sum_i \langle j |\hat{\sigma_z}_i| j \rangle$ are the sum of the expectation values of the operator $\sigma_z$ for each spin state $| j \rangle$.
The evolution of these expectation values is ruled by the following set of equations 
\begin{eqnarray} \nonumber
\frac{\partial S}{\partial t}&=&\big[-\frac{1}{T_2^*}  + i(\Delta+\gamma_e B \phi) \big]S  +i\frac{\Omega}{2} (S_z^{-1}-S_z^{0})\\ \nonumber
\frac{\partial S_z^{+1}}{\partial t}&=&-\frac{1}{T_1} (S_z^{+1}-S_z^{0})-\gamma_{\rm las} S_z^{+1}\\ \nonumber
\frac{\partial S_z^{-1}}{\partial t}&=&-\frac{1}{T_1} (S_z^{-1}-S_z^{0})-\gamma_{\rm las} S_z^{-1}
+i\frac{\Omega}{2} (S-S^*)\\ \nonumber
\frac{\partial S_z^{0}}{\partial t}&=&- \frac{\partial S_z^{-1}}{\partial t}- \frac{\partial S_z^{+1}}{\partial t}.\label{BE1}
\end{eqnarray}  
The last equation ensures that the spin population in this manifold is preserved. 
The important ingredient here is the angular dependent detuning that appears in the first equation. 
These coupled equations of motion are then coupled to the angular motion {\it via}
\begin{eqnarray} 
\frac{\partial^2 \phi}{\partial t^2}&=& -\omega_\phi^2 \phi - \gamma \frac{\partial \phi}{\partial t}+\Gamma (S_z^{-1}-S_z^{+1})+\Gamma_T(t)
\end{eqnarray} 
where $\Gamma=\hbar N \gamma_e B/I$. This equation describes the angle evolution under the NV spin torque.
We note that if the spin population in the two magnetic excited states is the same, no torque is applied to the diamond. 
$\Gamma_T(t)$ term is the delta-correlated Langevin noise term. 

In the experiment, the polarization dynamics and librational motion evolve on $ms$ timescales, much longer than the $T_2^*$ time ($ns$). We can thus adiabatically eliminate the evolution of $S$ and use rate equations that describe the spin populations only.
They read 
\begin{eqnarray} 
\frac{\partial S_z^{+1}}{\partial t}&=&-\frac{1}{T_1} (S_z^{+1}-S_z^{0})-\gamma_{\rm las} S_z^{+1}\\ \nonumber
\frac{\partial S_z^{-1}}{\partial t}&=&(-\frac{1}{T_1}+\Omega^2 P(\Delta,\phi)) (S_z^{-1}-S_z^{0})-\gamma_{\rm las} S_z^{-1}\\ 
\frac{\partial S_z^{0}}{\partial t}&=&- \frac{\partial S_z^{-1}}{\partial t}- \frac{\partial S_z^{+1}}{\partial t}
\end{eqnarray} 
where $P(\Delta,\phi)$ is given by 
\begin{eqnarray} 
P(\Delta,\phi)&=& \frac{1}{2\sigma}\frac{1}{1+(\frac{\Delta+\gamma_e B\phi}{\sigma})^2}
\end{eqnarray} 
where $\sigma=1/T_2^*$.
$P(\Delta,\phi)$ is a Lorentzian function quantifying the change in the polarization rate of the NV centers in the magnetic state. Using this model, the spin resonance lineshape would not accurately model our experiment, where the coupling of the NV centers to a paramagnetic spin bath gives rise to a Gaussian lineshape. Here, since the spin bath dynamics is fast compared the measurement time, it does not play any other role than changing the actual lineshape, we can thus instead write $P$ as a Gaussian function 
\begin{eqnarray} 
P(\Delta,\phi)&=& \frac{1}{2\sigma} \exp(-\frac{(\Delta+\gamma_e B (\phi_0+\phi))^2}{\sigma^2})
\end{eqnarray} 
with a width given by $é\sigma \sqrt{{\rm ln}(2)}/2\pi$.

The solutions for $\phi$ obtained from this numerical model can be used to estimate the effective damping and spin-spring effects and hence the temperature. 

\subsubsection{Relation between the spin population fluctuations and the final temperature}

%The equation for $\phi$ is  
%\begin{eqnarray}I \ddot \phi&=& -I \omega_\phi^2 \phi -I\gamma \dot \phi+\Gamma_T+\Gamma_{\rm mag}
%\end{eqnarray}\label{eqnmotion}
%Where the magnetic torque from the NV spins at a position $\pi/4$ is given by
%$$\Gamma_{\rm mag}= I\Gamma (S_z^{-1}-S_z^{+1})$$

To understand how to relate the results from our numerical simulations to the effective temperature, we can again write the equation of motion for $\phi$ in the Fourier domain 
\begin{eqnarray}
\phi(\omega)&=&\chi(\omega)(\Gamma_T(\omega)+\Gamma_{\rm mag}(\omega))\label{susc}
\end{eqnarray}
%\begin{figure}[ht!]
%\centerline{\scalebox{0.4}{\includegraphics{figbib_SI/Cooling_spring.pdf}}}
%\caption{a) b) Ring down measured by suddenly turning on a microwave signal tuned to the blue or the red of the NV transition b) Evolution of the spin population, angle, effective damping and frequency as a function of the microwave detuning from the full numerical simulations.}\label{Cooling}
%\end{figure}
As was done in \cite{SI_ArcizetPRA}, in the linear regime, one can also introduce an effective susceptibility $\chi_{\rm eff}$ such that
$$\phi(\omega)=\chi_{\rm eff}\Gamma_T(\omega).$$
%$$\chi(\omega)=\frac{1}{I(\omega^2-\omega_\phi^2+i\omega \gamma)}$$
In order to find $\chi_{\rm eff}$, one needs a relationship between the angle and the spin population.
To do this, we decompose the spin and angles as the sum of a mean value and a fluctuating component, as $S_z=\overline{S_z}+\delta S_z$ for instance. 
Linearizing the Bloch equations will then allow to get a relationship between $\delta S_z$ and $\phi$ 
$$\delta S_z=\xi(\omega) \phi.$$
where $\xi$ is a complex number that depends on all the parameters of the spin-mechanical interaction. 
Injecting this relation back into Eq. (\ref{susc}), we get 
$$\phi=\chi(\omega) \delta \Gamma_T + \chi \hbar \gamma_e B N \xi(\omega) \phi.$$
%This is a self-referenced relationship, characteristic of induction phenomena.
The power spectral density of the librational motion $\phi$ can finally be found to be 
\begin{eqnarray}
S_\phi(\omega)&=&\Big | \frac{\chi(\omega)}{1-  \hbar \gamma_e B N \xi \chi(\omega)}   \Big |^2 S_T(\omega)\\
&=&|\chi_{\rm eff} |^2 S_T(\omega).
\end{eqnarray}

It was shown in \cite{SI_ArcizetPRA}, that the real part of $\xi$ gives either an extra binding or anti-binding confinement (depending on its sign), also know as a spring effect. 
The imaginary part of $\xi$, is related to delayed action of the spin onto the angle and provides the cooling/heating mechanism.
Provided that the dynamics is still that of a damped harmonic oscillator, effective damping and spring effects can be found. 

We checked this by solving the full set of numerical simulations. 
The set of experiments shown in Fig. 2-d) where performed by exciting parametrically the librational motion using the internal spins inside the diamond. The ring down of the librational mode was then observed and was very well approximated by exponentially decaying curves (see Methods).

For the most part of the paper, we model the experiment numerically using Monte-Carlo simulations and the XMDS package \cite{SI_DENNIS2013201}. 
For the spin-cooling and spin-spring effects shown in Fig. 2.c), we used the measured librational frequency $\omega_\phi=2\pi\times$ 480 Hz, the measured damping rate $\gamma/2\pi=16$ Hz and the magnetic field splitting $\gamma_e B=260$ MHz as parameters.
The number of NV centers, polarization rate, Rabi frequency and the longitudinal spin lifetime $T_1$ in the excited state are not known with a high precision and are left as free parameters.
The number of NV centers in particular can be deduced from the detected count rate, but this can only be an order of magnitude estimate since we do not know precisely the collection efficiency of the apparatus, how the laser light is coupled to the NV centers in the diamond and the PL is refracted by the diamond. Doing this we found around $10^9$ NV centers in total, but use it as a free parameter when determining the best value in the experiment. Similarly, the moment of inertia is not known precisely due to the imprecision in the quoted diamond diameter $d$ and shape, which translates into a large error in the moment of inertia (which scales like $d^5$).
We thus set the parameter $\Gamma$ as a free parameter. 
%Figure \ref{Cooling}-a) show the result of numerical simulations where a microwave signal is tuned to the blue or the red of 
%the transition, and applied at a time t=20 ms, using the following parameters expressed in kHz :
Numerically, we turn on a microwave at a time t=20 ms, and record the librational mode ring down.
When the microwave is quasi-resonant, it induces a spin torque which displaces the angle from the equilibrium position. 
To find the optimum values for the resulting cooling and spring-constants, and to obtain a fit to the experimental data, we analyze $S_z$, $\theta$, $\gamma_{\rm eff}$ and 
$\omega_{\rm eff}$ as a function of the microwave detuning from the spin transition. 
%The results of such a numerical scan is shown in Fig. \ref{Cooling}-b). 
The numerical results of the Monte-Carlo simulation are adjusted to match the experimental parameters corresponding to the Fig.2-c) of the paper. Doing a Monte-Carlo simulation was important to describe damping of the oscillations on the blue-side. 
Without the Brownian motion of the angle included in the numerical simulations, strong phonon-lasing (described below) takes place in the regime where the damping on the red matches the experimental data. The Brownian motion induces phase diffusion of the laser, which effectively damps out our averaged signal. 

Using these numerical data, we can then fit a cosine function multiplied by an exponentially decaying curve to deduce both the spin spring and damping rates.  We observe an increase/decrease in the trapping frequency and a decrease/increase of the damping rate when the microwave is tuned to the blue/red. The analytical formula is in very good agreement with the ring down numerical data with $1/T_1=0.6$, $\gamma_{\rm las}=2$ and $\Omega/2\pi=30$ in kHz units, as free parameters.

We can then deduce effective damping and spring coefficients when the microwave is tuned to the red side of the spin resonance, and write the net susceptibility of the levitating diamond angle to the spin-torque as 
$$\chi_{\rm eff}(\omega)=\frac{1}{I(\omega_{\rm eff}^2-\omega^2+i\omega \gamma_{\rm eff})}.$$
from which we can deduce the final temperature using the equipartition theorem.

The imaginary part of $\xi(\omega)$ determines the damping term in the mechanical susceptibility, so 
the delay between the librational motion and the depolarization is the cause of cooling/heating. 
This model reproduces qualitatively the experimental results. 
To optimize the cooling, the laser induced re-polarization rate should thus be on the order of the trapping frequency.
Again, this simplified analysis neglects the $T_1$ decay, which tends to demagnetize the whole spin ensemble on millisecond times scales.
The librational frequency must thus be on the order of the $T_1$ time, which is the main limitation to the cooling efficiency in the experiment. 
Increasing the microwave and laser powers makes the librational mode evolution become bistable when the microwave is tuned to the red, and motional lasing occurs on the blue as we describe next.

%Operating with a highly confining trap was mandatory to reach a strong cooling. This may explain the differing damping rates observed for the two different librational modes of Fig. 2-c).

 \subsection{The non-linear regime : Bistability and phonon-lasing}
 
% \begin{figure}[ht!]
%\centerline{\scalebox{0.15}{\includegraphics{figbib_SI/bistable_scan.pdf}}}
%\caption{a) Effective spin-mechanical potential and b) the corresponding real solutions of the angular motion in the bistable regime as a function of the microwave detuning. }\label{bistable}
%\end{figure}

The above calculations were performed in the regime where the spins were weakly polarized by the laser and microwaves.
%Increasing the amplitude of the microwave power has different effects when it it tuned to the blue or to the red 
%so obtaining a general condition for the applicability of the linear approximation is not trivial. 
Here, we use the simplified two-level model to express the effective spin-mechanical potential energy in the non-linear regime. 
We will explain Fig. 3 of the main text, where bistability and phonon-lasing of the librational modes are observed. 
Let us first discuss the regime where the microwave is tuned to the red. 

\subsubsection{Bistable regime}

The phenomenon of bistability has been analyzed and experimentally realized by many research groups and can be observed in a vast range of experimental settings.  Bistability is associated with a wealth of interesting phenomena \cite{SI_LUGIATO198469}, one prominent example being the possibility to observe driven-dissipative phase transitions \cite{SI_Minganti}. One situation that has been extensively studied is when atoms are placed in an optical cavity, which was realized in the early days of cavity Quantum electrodynamics \cite{SI_RempeG}. 
The present system bears strong analogy with these experiments, since the NV spin degree of freedom plays the role of the cavity field in the experiments.

In order to describe this observed bistability, we can start from a simple two-level model. 
In the steady state limit, the Bloch equations give the following equation for $S_z$ 
$$S_z=\frac{2\Omega^2}{\gamma_{\rm las} T_2}\Big[\frac{1}{(1/T_2^*)^2+(\Delta+\gamma_e B \phi)^2+4\Omega^2/ (\gamma_{\rm las} T_2^*) }\Big].$$
Using the equation of motion for the angle in the steady state limit gives 
$$\phi=I \Big[\frac{1}{\kappa+(\Delta+\gamma_e B \phi)^2}\Big],$$
where $\kappa=(1/T_2^*)^2+4\Omega^2/ (\gamma_{\rm las} T_2^*) $ and $I=\frac{\Gamma}{\omega_\phi^2}\frac{2\Omega^2}{\gamma_{\rm las} T_2}$. This relation is a third order polynomial equation for $\phi$. We solve it numerically using the parameters of our experiment and found that when $\Delta<0$ it can have two real and one imaginary solutions. 
Physically, this effectively corresponds to a stable and unstable (saddle point) for the angle.  
The result of the numerical simulations is shown in Fig. 3-a) as a function of the detuning, where a characteristic S-shape curve is observed. 
When the microwave is tuned to the blue of the spin resonance transition, two stable solution for the angle (at positions A and B) are seen to co-exist.  
%Getting to these two angular positions depend on the history of the particle. 
%The particle being capable of hoping from site A to B via collisions with gas particles or other noise sources.
%And the particle angle in our experiment plays the role of the atomic energy in the cavity QED experiments.

The probability of being in either one of these two points depends on the history of the angle motion, which often implies a hysteretic behavior. In the experiment, we expect that if the microwave is scanned from the blue to the red,
the particle will follow the upper angular trajectory (where the saddle point is B) up to the end, and finally drop to the small angle solution. On the other hand, if the microwave is scanned from the red to the blue, the particle will stay in the saddle point A  much longer until the spin torque pulls it towards large angles. 
If the scan is performed on times scales of the librational period then hysteresis behavior is expected.
%The change in magnetisation would then due to the fact that the microwave is scanned at a rate that is on the order of the particle motion.
This hysteresis behavior was observed in the experiment (Fig.3-b)) and numerical simulations of the coupled Bloch and torque equation very well describe the data.

If now the microwave is parked to the red of the spin resonance, there may be sudden stochastic jumps between the two metastable angular positions. 
External noise sources may indeed push the angle from one point to the other so that sharp jumps from the two stable positions A and B may occur on times scales of the motional period, and remain at the stable points for times scales dictated by the amplitude of the external driving noise. The effect was observed in the experiment, and is shown in Fig. 3-c) of the main text. 
The origin of this jump is likely to be dominated by collisions with the gas particles, but any other sources of noise, such as the microwave signal generator amplitude noise or laser noise, could also explain our observations. 

We also measured the relative time spent in each of these stable dynamical potential wells in these condition (long measurement times) as a function of the microwave detuning. 
The evolution of the histograms showing the time spent in each of those wells is shown in the extended data Fig. \ref{bistamicro}.

%Assuming that the two stable points are separated by a potential with a barrier height $U(\Delta)$.
%Including the steady-state spin response to the microwave drive in the total energy, the energy reads
%$$E=\frac{1}{2}I\omega_\phi^2\phi^2-N\hbar(\gamma_e B\phi_0+\Delta)\overline{S}_z(\phi_0)$$
%Here, $\overline{S}_z$ depends on the particle angle $\phi_0$ due to the change in the resonance condition when the particle rotates.
%The total spin-mechanical potential energy is plotted in Joules, as a function of angle in Fig. 3-a) of the main text for the parameters of our experiment and using the two-level theory.
%We took $I=10^{-23} kg/m^2$, $N=5\times10^9$. 
%
These results point towards a naive interpretation that the two local potential minima are effectively separated by a potential barrier whose height depends upon the microwave detuning. 
%In this regime, bistability and strong hysteresis can occur, the particle being capable of hoping from site A to B via collisions with gas particles or other noise sources.
Going further in this interpretation, one could estimate that the jump rate between two local minima from Kramer's theory \cite{SI_Ricci, SI_RondinKramers}. In the overdamped damped regime, the transition rate from the well A to B is :
\begin{equation}
R_{\rm AB}=\frac{1}{2\pi} \prod_{i=x,y,z} \frac{\omega_i^A}{|\omega_i^C|}\Big[\sqrt{|\omega_i^C|^2+\gamma^2/4}-\gamma/2\Big]\exp(\frac{-U_A}{k_B T})
\end{equation}
where $\omega_i^{A,B,C}$ correspond to the angular confinement frequencies of the mode $i$ at the point A,B or C. 
The reverse rate is obtained by swapping indices A and B.
In our measurements the particle librational is underdamped
so the transfer is slowed down by the slow transfer of energy between the librational mode and the bath, but the prefactor for the transition rates from A to C and C to A is the same, up to a difference in the librational mode frequencies in A and B \cite{SI_Ricci, SI_RondinKramers}.

Here, we are interested in the ratio $R=P_A/(P_A+P_B)$, which quantifies the normalized averaged population in the site A. 
This ratio is mainly governed by the difference in the potential height and effective temperature, including all noise sources, so 
we assume that the two frequencies at the points A and B are of the same order of magnitude for simplicity. 
\begin{equation}
R=\exp(\frac{U_A(\Delta)-U_B(\Delta)}{k_B T})
\end{equation}
Such a dependence of $P_A$ with the temperature of the bath is also confirmed by the Monte-Carlo simulations 
and offers a rich playground to studies of thermalization in out-of-equilibrium systems.
%The potential height can be estimated from the shape of the hamiltonian in steady state using the simplified two-level model, which fits well our experiment well.

\subsubsection{Librational lasing}

Phonon-lasing takes place when stimulated emission overcomes absorption of phonons from the mechanical mode. 
This effect has recently been observed using the mechanical modes of micro-toroids in \cite{SI_Grudinin}.
Here the phonon laser is replenished via the dual pumping mechanism provided by the microwave (why magnetizes the NV states) and the green laser that polarises the NV in the ground state. In addition, delay also plays a crucial role in the phonon-lasing process. 
Indeed, the critical point where phonon-lasing occurs is to be found when the damping $\gamma_{\rm eff}$ becomes negative, that is when the delay makes the system unstable. 
Fig. 6-a) shows the result of numerical simulations in this regime. 
Numerical simulations show the damping and spin-spring shifts with a torque coefficient $\Gamma={0.7, 0.9, 1.2,1.5}$ kHz from traces i) to iv) and resonant microwave Rabi frequency of 100 kHz. 
On the blue side, the damping coefficient becomes negative and the spring effect stabilizes to a value that is slightly greater than the librational mode frequency defined solely by the Paul trap (here 480 Hz). This is observed experimentally in the Fig. 3.b)-i) of the main text. In this regime, the oscillator runs unstable and self sustained oscillations then take place.
In the transient regime, the librational mode undergoes amplification, until it settles to a constant coherent oscillation. 
This is shown Fig.~6-b), where a microwave signal detuned by 7 MHz from the spin resonance spin transition is applied at a time t=20 ms. Here we set $\Gamma$ to 1.5 kHz.
At this detuning, the signal is amplified for 120 ms, up to a point where a regime of self-sustained oscillation sets-in with an angular amplitude that settles at 0.05 rad. 

Phonon-lasing occurs at a threshold value that depends on the ratio between losses and gain. 
The losses are related to residual gas damping and the longitudinal spin relaxation time $T_1$ limits the population inversion. 
Numerical simulations of the librational mode velocity are plotted as a function of microwave Rabi frequency in Fig. 6-c) for three different values of the spin-lattice relaxation rate $1/T_1=\{1, 2, 3\}$kHz for trace i)-ii) and iii) respectively. 
A threshold behavior is observed at microwave Rabi frequencies of $24, 38$ and 59 kHz for the traces
i),ii) and iii).  This theory is in good agreement with the experimental observations shown in Fig. 3.b)-ii) in the main text.

\newpage

\end{widetext}

\end{document}